\documentclass{cargese}
\let\footnote\savefootnote
\let\footnotetext\savefootnotetext 
\let\footnoterule\savefootnoterule 
\normallatexbib

\begin{document}

\articletitle[Non-supersymmetric open-strings with a quantised 
$B_{ab}$]{A Note on non-supersymmetric
\\ open-string orbifolds with a \\ quantised
$B_{ab}$ ${}^*$\footnote{Preprint CPHT-PC231.0899, LPTENS 99/30}}


\author{Carlo Angelantonj}

\affil{Centre de Physique Th{\'e}orique,  {\'E}cole Polytechnique
\\ 
F-91128 Palaiseau Cedex
\\
and
\\
Laboratoire de Physique Th{\'e}orique de l'{\'E}cole Normale Sup{\'e}rieure
\\
24 rue Lhomond, F-75231 Paris Cedex 05}

\email{angelant@cpht.polytechnique.fr}

\begin{abstract}
In this short note we review the main features of open-string orbifolds
with a quantised flux for the NS-NS antisymmetric tensor in the
context of the open descendants of non-supersymmetric asymmetric orbifolds
with a vanishing cosmological constant.
\end{abstract}


Recently we have analysed in some detail the complete structure of the
one-loop amplitudes for open strings \cite{cargese,bs} on
orbifolds \cite{ps} with a
quantised NS-NS antisymmetric tensor background \cite{tororb}, thus
extending the results in \cite{toroidal}. 
Besides recovering the expected rank reduction for
the Chan-Paton (CP) gauge group, we were led to identify the discrete
Wilson lines of \cite{twist} with the signs $\gamma_\epsilon$ that
enforce a correct normalisation of the M{\"o}bius amplitude. 
In particular, for $Z_2$ orbifolds, they
allow one to connect continuously ${\rm U} (n)$ groups to ${\rm Sp}
(n) \otimes {\rm Sp} (n)$ groups.
In this brief note we want to reconsider the non-supersymmetric type I
vacua with vanishing cosmological constant discussed in
\cite{bg,nonsusy} in the spirit of \cite{tororb}. We refer to the original
papers for more details and for references.

Non-supersymmetric vacua with a vanishing cosmological constant can be
obtained as asymmetric orbifolds of type II and/or type I
superstrings. The simplest instance of these models can be obtained
considering the generators \cite{kks}
\begin{eqnarray}
f &=& [ (-1^4 , 1;1^5),(0^4 , v_{{\rm L}} ; \delta ^4 , v_{{\rm R}} ) ,
(-1)^{F_{{\rm R}}} ]\,, \nonumber
\\
g &=& [ (1^5 ; -1^4 ,1),(\delta ^4 , w_{{\rm L}} ; 0^4 , w_{{\rm R}} ) ,
(-1)^{F_{{\rm L}}} ]\,.
\label{gen}
\end{eqnarray}
The first entry in the square brackets denotes rotations, the
second denotes shifts on the internal compactification lattice while the third
corresponds to the space-time fermion-number. The shift $\delta$ acts as
a $Z_2$-shift, whereas $v_{{\rm L,R}}$ ($= w_{{\rm R,L}}$) act as an
$A_2$ shift \cite{vw} on the fifth coordinate. The asymmetric
nature of the orbifold together with level matching, require that the
internal 4d lattice split into a product of four circles at the
radius  of SU(2) enhanced symmetry. Actually, the algebra of the
generators (\ref{gen}) reveals that $f$ and $g$ generate a non-abelian 
space group orbifold $S$ \cite{dhvw}. The restriction to the point group 
$\overline{P} = S / \{f^2 , g^2\}$, with $f^2$ and $g^2$ pure
translations, effectively
changes the lattice to an SO(8) lattice, thus introducing a
non-vanishing background for the $B$-field 
$$
B = {\alpha ' \over 2} \left(\begin{array}{cccc} 0&-1&0&0\\
1 & 0 & -1 & -1 \\ 0 & 1 & 0 & 0 \\ 0 & 1 & 0 & 0 \end{array} \right)
$$
with rank $r=2$. 

We have
already noticed in \cite{nonsusy} how the choice of the internal
 SO(8) lattice is crucial in order to get a consistent
result. In fact, the orbifold that we are considering can be thought
as an $f$ (or $g$) projection of the supersymmetric $T^4/Z_2$ orbifold 
generated by $fg$, with the $T^4$ corresponding to the SO(8) lattice.
The resulting torus amplitude
$$
{\cal T}_0 \sim |V_4 O_4 |^2 + |S_4 S_4 |^2 - (O_4 V_4 )(\bar C_4 \bar
C_4 ) - (C_4 C_4 )(\bar O_4 \bar V_4) + \ldots
$$
is then consistent with the generic Klein bottle amplitude\footnote{In
our case the spinors have a
flipped chirality, since the supersymmetric $Z_2$
generator $fg$ contains $(-1)^{F_{{\rm L}} + F_{{\rm R}}}$ together
with the standard inversions.} (that, roughly speaking feels only
the left-right symmetric generator $fg$) 
\cite{tororb}
$$
\tilde{\cal K}_0 \sim \left[ \sqrt{v} + {2^{-r/2} \over \sqrt{v}}
\right]^2 (V_4 O_4 - S_4 S_4 ) +
\left[ \sqrt{v} - {2^{-r/2} \over \sqrt{v}}
\right]^2 (O_4 V_4 - C_4 C_4 )
$$
only for an internal SO(8) lattice, where the
volume $v={1\over 2}$ and $r=2$.
Thus $\tilde{\cal K}_0$ contains only those states that in
${\cal T}_0$ are combined with their antiholomorphic counterparts. 
The same
phenomenon also presents itself in the open sector, upon identification of
Neumann and Dirichlet CP charges as a result of modding out by the
T-duality contained in $f$. As a result of the action of $f$, the CP
gauge group is reduced further and turns into in a single U(8)
factor \cite{bg,nonsusy}. Actually, this is the case if we break the
residual internal global ${\rm SO} (4)^2$ symmetry by introducing
discrete Wilson lines in the M{\"o}bius amplitude, thus affecting the
$P = T^{{1\over 2}} S T^2 S T^{{1\over 2}}$ transformation, that 
acts on the real hatted characters
\cite{twist,tororb}. Let us analyse the massless
contribution to the transverse-channel M{\"o}bius amplitude:
$$
\tilde{\cal M}_0 \sim -{\textstyle{2\over 2}} (M_1+ M_2 ) [ (\hat V_4
\hat O_4 - \hat S_4 \hat S_4 ) \hat O_4 \hat O_4 - (\hat O_4 \hat V_4
- \hat C_4 \hat C_4 ) \hat V_4 \hat V_4 ] \,,
$$
where with have explicitly written the contribution of the internal
bosons. On the SO(4) (hatted) characters $(\hat O_4 , \hat V_4 , \hat
S_4 , \hat C_4)$, the $P$ matrix acts as ${\rm diag} (\sigma_1 ,
\sigma_1)$, with $\sigma_1$ the first Pauli matrix. As a result,
the space-time vector contributes to
$$
{\cal M}_0 \sim {\textstyle{1\over 2}} (M_1 + M_2 ) [ (\hat V_4
\hat O_4 - \hat S_4 \hat S_4 ) \hat O_4 \hat O_4 - (\hat O_4 \hat V_4
- \hat C_4 \hat C_4 ) \hat V_4 \hat V_4 ]
$$
thus calling for a product of symplectic gauge groups 
${\rm Sp}(8) \otimes {\rm
Sp} (8)$, consistently with the tadpole conditions. The resulting model
is still supersymmetric at all mass levels in the open sector, while
it is non-supersymmetric in the closed sector. This has to be
contrasted with the recently proposed scenario \cite{bulk,tororb} where
supersymmetry is unbroken in the bulk (at any mass level) but is
broken on branes, where a positive cosmological constant is generated at one
loop. 

The U(8) model of \cite{bg,nonsusy} is obtained introducing
in  $\tilde{\cal M}$ discrete Wilson lines, that result in
primed SO(4) characters associated to the
internal lattice \cite{twist,tororb}:
\begin{eqnarray}
\hat O _4 = \hat O_2 \hat O _2 - \hat V_2 \hat V_2 &\to & \hat O ' _4
= \hat O_2 \hat O_2 + \hat V_2 \hat V_2 \,,
\nonumber
\\
\hat V _4 = \hat O_2 \hat V _2 + \hat V_2 \hat O_2 &\to & \hat V ' _4
= \hat O_2 \hat V_2 - \hat V_2 \hat O_2 \,,
\nonumber
\end{eqnarray}
whose $P$ transformation is now given by the Pauli matrix $\sigma_3$
acting on $(\hat O '_4 , \hat V ' _4)$. As a result
$$
{\cal M}_0 \sim -{\textstyle{1\over 2}} (M + \bar M ) [ (\hat O_4 \hat
V _4 - \hat C_4 \hat C_4 ) \hat O '_4 \hat O '_4 - (\hat V_4 \hat O_4
- \hat S _4 \hat S_4 ) \hat V '_4 \hat V'_4 ] \,,
$$
consistently with a unitary CP gauge group. 

Although this latter choice has a sensible ``geometric'' 6d
decompactification limit, it is not naturally related to the 
rational SO(8) internal lattice.
In fact, it is the ${\rm Sp} (8) \otimes {\rm Sp} (8)$ model 
that, in the 6d decompactification limit, leads to 
the ${\rm Sp}(8)^4$ model of \cite{bs,twist,gepner} pertaining to the
rational SO(8) lattice.


\begin{acknowledgments}
It is a pleasure to thank I. Antoniadis, K. F\"orger and A. Sagnotti for
interesting discussions. I would also like to thank
the Organizers of the Carg{\`e}se Summer School for having created a
pleasant and stimulating atmosphere.
This work was supported in part by EEC TMR contract 
ERBFMRX-CT96-0090.
\end{acknowledgments}

\end{document}